# A complementary media invisibility cloak that can cloak objects at a distance outside the cloaking shell

Yun Lai, Huanyang Chen, Zhao-Qing Zhang and C. T. Chan

Department of Physics
The Hong Kong University of Science and Technology
Clear Water Bay, Kowloon, Hong Kong, China

## Abstract

Based on the concept of complementary media, we propose an invisibility cloak operating at a finite frequency that can cloak an object with a pre-specified shape and size within a certain distance outside the shell. The cloak comprises of a dielectric core, and an "anti-object" embedded inside a negative index shell. The cloaked object is not blinded by the cloaking shell since it lies outside the cloak. Full-wave simulations in two dimensions have been performed to verify the cloaking effect.

Recently, great progress has been made in both the theory and experiment of invisibility cloak [1-18]. One approach to achieve the invisibility cloak is to employ the transformation optics concept to exclude electromagnetic waves in certain regions and smoothly fit them to fields outside the device [1-13]. The permittivity and permeability of such a cloak are determined by the coordinate transformation [1, 2] of expanding a point or a line into a finite volume of cloaked space. Such a transformation media cloaking device can hide any object inside the cloaked domain. The object hidden inside the cloaked domain has to be "blind", since no outside electromagnetic waves can reach into the cloaked space. Another approach of cloaking is to reduce the lowest order scattering cross section of an object by covering it with layers of metamaterials [14, 15]. In this approach, the properties of the cloaking device depend on the object to be cloaked. In the approaches mentioned above, the cloaking shell encloses the object to be hidden. It is also possible to have cloaking effects on an object which lies outside the cloaking shell. In particular, it has been mathematically proven that a polarizable line dipole in a matrix of dielectric constant $\varepsilon_m$ becomes invisible in the vicinity of a coated cylinder with core dielectric constant $\varepsilon_c \approx \varepsilon_m$ and surface coating dielectric constant $\varepsilon_s \approx -\varepsilon_m$ in the quasistatic limit [16-18]. The origin of this "external" cloaking effect is the anomalous localized resonance of the coated cylinder that cancels any induced moment.

In this paper, we propose a new recipe of an invisibility cloak that can hide an object that is external to the cloak itself. The cloak depends on the object, but the object can have arbitrary shape. The idea is based on combining the concept of complementary media and transformation optics [19-23]. It is known that complementary media can optically "cancel" a certain volume of space at a certain frequency. This concept has important implications, most notably in the "perfect lens" [19] and novel imaging devices [20], and has motivated various applications, such as the "superscatterer" [24, 25], the cylindrical superlens [26], the "anti-cloak" [27], and an alternative strategy of invisibility cloak (which is different from our approach here) [13]. Complementary media of electrons have also been discussed [21]. Here, we show that a new type of invisibility cloak that employs an "anti-object" embedded in a negative index shell can make an object that lies outside the cloaking shell invisible. The working principle can be described in two steps. First, the object as well as the surrounding space is optically canceled by using a complementary media layer with an embedded complementary "image" of the object, which is referred as the "anti-object" hereafter. Then, the correct optical path in the canceled space is restored by a dielectric core material. As a result, the total system is effectively equal to a piece of empty space fitted into the cancelled space, and invisibility is achieved.

The key idea behind this design is the complementary media [19-22] which can be regarded as a special kind of transformation media. According to the coordinate transformation theory, when a space is transformed into another space of different shape and size, the permittivity $\boldsymbol{\varepsilon}'$ and permeability $\boldsymbol{\mu}'$ in the transformed space $\mathbf{x}'$ are given by $\boldsymbol{\varepsilon}' = \mathbf{A}\boldsymbol{\varepsilon}\mathbf{A}^{\mathrm{T}}/\det\mathbf{A}$ and $\boldsymbol{\mu}' = \mathbf{A}\boldsymbol{\mu}\mathbf{A}^{\mathrm{T}}/\det\mathbf{A}$, where $\boldsymbol{\varepsilon}$ and $\boldsymbol{\mu}$ are the permittivity and permeability in the original space $\mathbf{x}$, and $\mathbf{A}$ is the Jacobian transformation tensor with components $A_{ij} = \partial x'_i/\partial x_j$. The transformation media exhibits two interesting properties. First, the optical path in the transformation media is exactly the same as that in the original space. Second, the transformation media is reflectionless as long as the outer boundary coordinates before and after a coordinate transformation is unchanged [22, 23]. Based on these principles, a kind of complementary media can be designed under a special kind of coordinate transformation, i.e. folding a piece of space into another. When a wave crosses the folding line from the original space into the folded space, it starts to experience a "negative" optical path as that in the original space. A good example of the complementary media is the perfect lens [19], as is shown in Fig. 1(a). A perfect lens is a slab of $\varepsilon' = -1$ and $\mu' = -1$, formed under a coordinate transformation of $x' = -x$ for $0 < x < L$, i.e. folding a slab of $0 < x < L$ into the slab of $-L < x' < 0$. Here $L$ is the slab width. In this case, the optical phase at $x = L$ and $x' = -L$ are exactly the same, and the region $(-L, L)$ appears to be nonexistent.

Now consider the simple scheme as shown in Fig. 1(b). An object with permittivity $\varepsilon_o$ and permeability $\mu_o$ is placed at the right (in a layer of air). A slab of complementary media can still be designed by the coordinate transformation $x' = -x$ for $0 < x < L$, which results in a slab of $\varepsilon' = -1$ and $\mu' = -1$ with an embedded complementary "image" object with permittivity $-\varepsilon_o$ and permeability $-\mu_o$. Then, similar to the case of perfect lens, the optical phase at $\pm L$ are exactly the same. The object and complementary media are optically canceled by each other. Plane wave will pass through without scattering.

Since a slab of complementary media is required to be infinitely long in the $y$ axis, it would be interesting to construct complementary media with a finite volume. In Fig. 1(c), we show the scheme of a circular layer of complementary media with parameters $\boldsymbol{\varepsilon}'$ and $\boldsymbol{\mu}'$ that optically cancels an outer circular layer of air. The complementary media can be obtained by a coordinate transformation of folding the layer of air ($b < r < c$) into the layer of complementary media ($a < r' < b$). Here $a$, $b$ and $c$ are the core radius, the outer radius of complementary layer, and the outer radius of the canceled air layer, respectively. Consider a general coordinate transformation of the form $r = f(r')$ [23], in which $f(r')$ is a continuous function of $r'$ that satisfies $f(b) = b$ and $f(a) = c$. The $\boldsymbol{\varepsilon}'$ and $\boldsymbol{\mu}'$ of the complementary layer can be written as $\varepsilon'_r = \mu'_r = \frac{f(r')}{r'} \frac{1}{f'(r')}$, $\varepsilon'_\theta = \mu'_\theta = \frac{r'}{f(r')} f'(r')$ and $\varepsilon'_z = \mu'_z = \frac{f(r')}{r'} f'(r')$. For a simple linear function $f(r') = (r' - b) \cdot (c - b)/(a - b) + b$, $\boldsymbol{\varepsilon}'$ and $\boldsymbol{\mu}'$ are anisotropic. But for the specific choice of $f(r') = b^2/r'$ and $c = b^2/a$, we obtain isotropic parameters $\boldsymbol{\varepsilon}'$ and $\boldsymbol{\mu}'$: $\mu'_r = \mu'_\theta = -1, \varepsilon'_z = -b^4/r'^4$ for TE waves, and $\varepsilon'_r = \varepsilon'_\theta = -1, \mu'_z = -b^4/r'^4$ for TM waves. We note in particular that the radial and tangential components are constants. If we let the core material to be a perfect electric conductor (PEC), the system becomes the so-called "superscatterer" [24, 25] with an effective PEC boundary at radius $c$. Here, in order to restore the optical path, we consider a core material with $\mu''_r = \mu''_\theta = \varepsilon''_z \cdot a^2/c^2 = 1$ for TE waves and $\varepsilon''_r = \varepsilon''_\theta = \mu''_z \cdot a^2/c^2 = 1$ for TM waves, which are obtained by coordinate transformation of $r = c/a \cdot r''$, i.e. compressing a large circle of air with radius $c$ into a small circle with radius $a$. With such a choice of dielectric core material, the wave experiences the same optical path as that in an circle of air with a radius $c$. In this way, the whole system, including the outer air layer, the complementary media layer and the core material, is optically equal to a circle of air with radius $c$, and is thus invisible together to any form of external illumination.

Now we go one step further. Suppose an object of permittivity $\boldsymbol{\varepsilon}_o$ and permeability $\boldsymbol{\mu}_o$ is added in the outer circular layer of air, and we want to make it invisible. Then, according to the transformation media theory, it is always possible to include a complementary "image" object with parameters $\boldsymbol{\varepsilon}'_o$ and $\boldsymbol{\mu}'_o$: $\frac{\varepsilon'_{or}}{\varepsilon_{or}} = \frac{\mu'_{or}}{\mu_{or}} = \frac{f(r')}{r'}\frac{1}{f'(r')}$, $\frac{\varepsilon'_{o\theta}}{\varepsilon_{o\theta}} = \frac{\mu'_{o\theta}}{\mu_{o\theta}} = \frac{r'}{f(r')}f'(r')$ and $\frac{\varepsilon'_{oz}}{\varepsilon_{oz}} = \frac{\mu'_{oz}}{\mu_{oz}} = \frac{f(r')}{r'}f'(r')$ in the complementary media layer, such that the object of $\boldsymbol{\varepsilon}_o$ and $\boldsymbol{\mu}_o$ is optically cancelled, as shown in Fig. 1(d). Thus, the object also becomes invisible to any incident waves. This recipe opens up a new way to conceal an object from electromagnetic waves within a specific distance outside the cloak.

In the following, we carry out full wave simulations using a finite element solver (Comsol Multiphysics) to demonstrate the functionality of the "external complementary media" invisibility cloaks. We consider cases of TE polarization (E along z-direction), and impose an incident plane wave from the left or a point souce with wavelength $\lambda = 1$ unit. First, we demonstrate the scheme shown in Fig. 1(c), i.e. a circular layer of complementary media ($0.5 < r' < 1$) and a core material ($r'' < 0.5$) that is optically equal to a circle of air ($r < 2$). Under a linear transformation mapping $f(r') = 3 - 2r'$, we obtain the parameters of the complementary media: $\mu'_r = (r' - 1.5)/r' \in [-2, -0.5]$, $\mu'_\theta = r'/(r' - 1.5) \in [-2, -0.5]$ and $\varepsilon'_z = (4r' - 6)/r' \in [-8, -2]$. In Fig. 2(a), the calculated electric fields are shown. The absence of scattered waves clearly verifies the invisibility of the whole system. Next, we demonstrate the scheme shown in Fig. 1(d), the cloaking of an object by complementary media. A dielectric curved sheet of thickness 0.3 with parameters $\varepsilon_o = 2$ and $\mu_o = 1$ to be cloaked is positioned between the circles of $r = 1.5$ and $r = 1.8$. In Fig. 2(b), the scattering pattern of such a single dielectric curved sheet is shown. In order to make the object invisible, we modify the complementary media layer in Fig. 2(a) to include a complementary "image" object, i.e. the "anti-object", with parameters $\varepsilon'_{oz} = 2\varepsilon'_z$ and $\boldsymbol{\mu}'_o = \boldsymbol{\mu}'$, positioned between the circles of $r' = 0.6$ and $r' = 0.75$. The cloak is composed of the modified complementary layer embedded with the "anti-object" and a core material. In Fig. 2(c), we show the calculated electric fields which clearly demonstrate the "external" cloaking effect. We note that the invisibility cloak does not cover the object surface here, and the cloaking effect comes from the optical "anti-object" embedded in the negative index shell. The scattering effect of the object and the "anti-object" cancel each other. We emphasize that there is no shape or size constraint on the object to be cloaked, as long as it fits into the region bounded between b<r<c. In Fig. 2(d), we demonstrate the cloaking of two curved sheets. The sheet on the left bounded between $1.2 < r < 1.5$ has an anisotropic permeability of $\mu_{or} = 1$ and $\mu_{o\theta} = -1$. The sheet on the downside bounded between $1.5 < r < 1.8$ has a linearly changing permittivity of $\varepsilon_o = 1 + (1.8 - r)/0.3$. In this case, an "anti-object" of $\mu'_{or} = \mu'_r$, $\mu'_{o\theta} = -\mu'_\theta$, $\varepsilon'_{oz} = \varepsilon'_z$ and an "anti-object" of $\varepsilon'_{oz} = \varepsilon'_z \cdot (1 - (0.6 - r')/0.15)$ and $\boldsymbol{\mu}'_o = \boldsymbol{\mu}'$, are embedded at the corresponding "image" positions in the negative index shell. Perfect cloaking effect is demonstrated with a point source positioned at (2,-2).

We will illustrate the cloaking scheme with another example. We will cloak a dielectric shell of $\varepsilon_o = 2$ and $\mu_o = 1$ bounded between $1.5 < r < 1.8$, as shown in Fig. 3(a), which also shows its scattering pattern under a plane wave illumination. In this case, the "anti-object" is an "image" shell with $\varepsilon'_{oz} = 2\varepsilon'_z$ and $\boldsymbol{\mu}'_o = \boldsymbol{\mu}'$, positioned between the circles of $r' = 0.6$ and $r' = 0.75$ inside the negative index complementary shell. In Fig. 3(b), we show the calculated electric fields after the dielectric shell of $1.5 < r < 1.8$ is "cloaked" by a shell composed of a complementary media $0.5 < r' < 1$ with "anti-object" and a core material $r'' < 0.5$. Please note again that the dielectric shell is outside the cloak. The perfect plane wave pattern manifests the cloaking effect. In Fig. 3(c), we consider another circular shell of $\varepsilon_o = -1$ and $\mu_o = 1$. The scattering pattern for such a shell, which is shown in Fig. 3(c) is similar to that of a metal shell. In this case, the "anti-object" is the complementary "image" shell with $\varepsilon'_{oz} = -\varepsilon'_z$ and $\boldsymbol{\mu}'_o = \boldsymbol{\mu}'$. In Fig. 3(d), the calculated electric fields after cloaking are shown. Again, an excellent cloaking effect is manifested by the perfect plane wave pattern.

We would like to compare this "anti-object" cloak to resonance cloak of Milton et al. While both can cloak an object external to the cloaking shell, Milton's cloak cloaks a point-like object (cross section is non-zero for finite-sized object) in the electrostatic limit. Our design works for finite size objects in a finite frequency.

In conclusion, we propose a perfect invisibility cloak that can make an object outside its domain invisible. The crucible element is a complementary "anti-object" embedded inside a shell with "folded geometry". This is different from the original perfect cloak [2], in which the hidden object is blind since it sits inside the cloaking shell. Our object sits outside the device, at the expense that the cloak must be "custom-made" as it has to carry an "anti-object". This scheme can be extended to three dimensions, and is applicable to any object that can be described by a given permittivity $\varepsilon_o$ and permeability $\mu_o$.

This work was supported by the Central Allocation Grant from the Hong Kong RGC through HKUST3/06C. We thank Dr. JunJun Xiao, DeZhuan Han, Jack Ng, KinHung Fung, ZhiHong Hang and Jeffrey ChiWai Lee for helpful discussions.

Phys. (2008), arXiv:0808.0536.

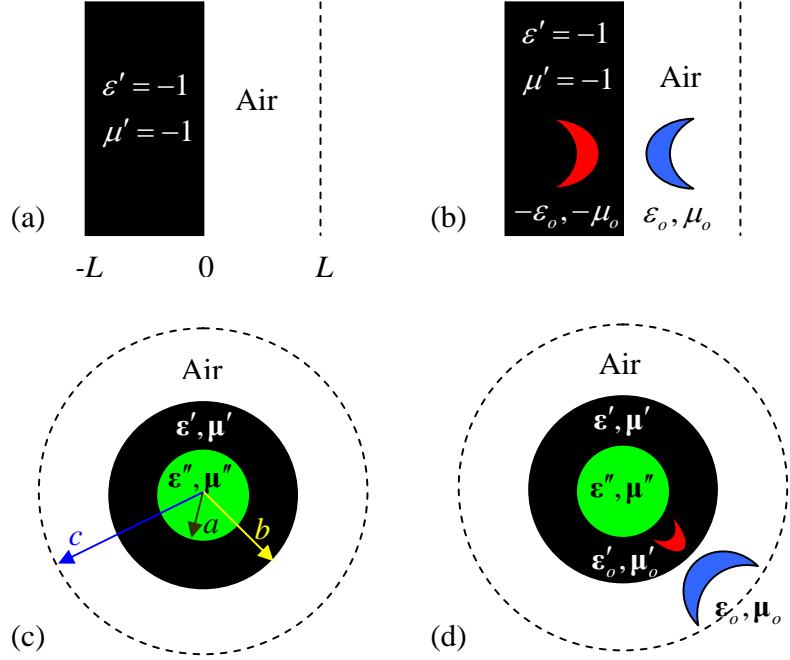

Fig. 1 (color online). (a) A complementary media slab $-L < x' < 0$ that optically cancels a slab of air $0 < x < L$. (b) A complementary media slab with an embedded complementary "image" of $-\varepsilon_o$ and $-\mu_o$ that optically cancels an object of $\varepsilon_o$ and $\mu_o$ in air. (c) A system composed of a circular layer of air ($b < r < c$), a circular layer of complementary media of $\boldsymbol{\varepsilon}', \boldsymbol{\mu}'$ ($a < r' < b$) and a core material of $\boldsymbol{\varepsilon}'', \boldsymbol{\mu}''$ ($r'' < a$) that is optically equal to a large circle of air ($r < c$). (d) A scheme to cloak an object of $\boldsymbol{\varepsilon}_o, \boldsymbol{\mu}_o$ by placing a complementary "image" of the object, i.e. the "anti-object" of $\boldsymbol{\varepsilon}'_o, \boldsymbol{\mu}'_o$, in the complementary media layer of $a < r' < b$.

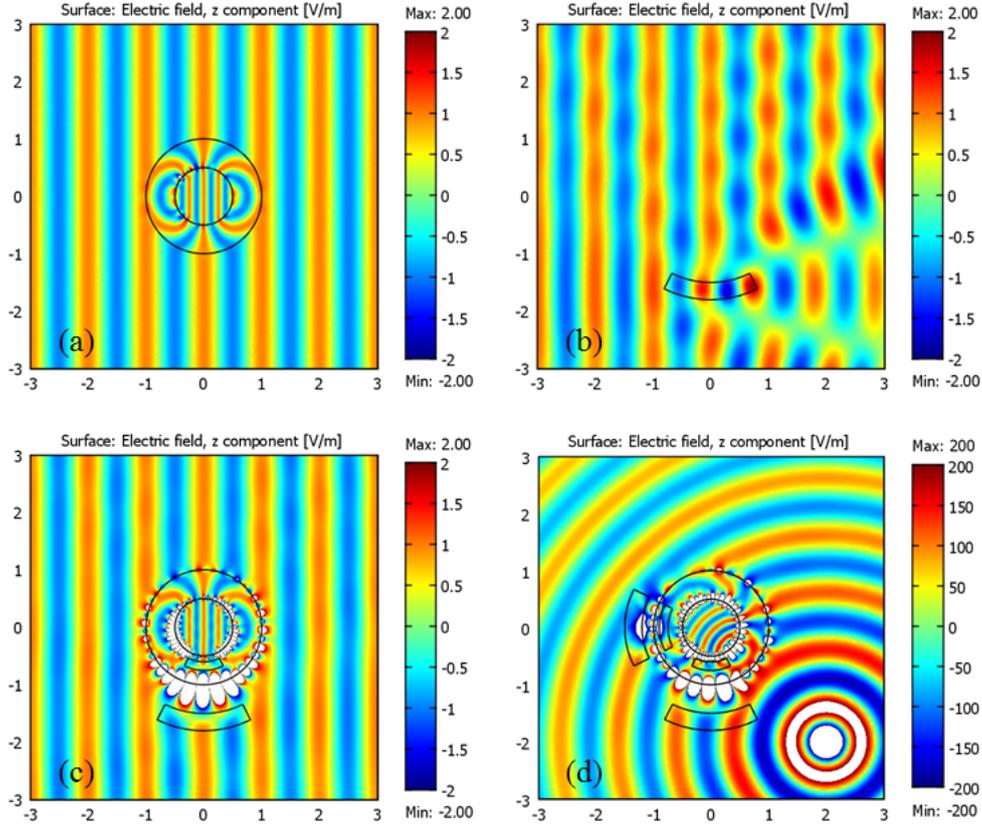

Fig. 2 (color online). Snapshots of the total electric fields under an incident TE plane wave from the left ((a)-(c)) and a point source ((d)). (a) A circular complementary media layer of $0.5 < r < 1$ and a core material of $r < 0.5$. (b) A curved sheet of thickness 0.3, with permittivity $\varepsilon_o = 2$. (c) The curved sheet in (b) is cloaked by an invisibility cloak composed of a circular layer of complementary media with an embedded "anti-object" and a core material. (d) A curved sheet on the left with anisotropic permeability $\mu_{or} = 1$ and $\mu_{o\theta} = -1$, and another curved sheet on the downside with permittivity $\varepsilon_o = 1 + (1.8 - r)/0.3$, are both cloaked by an invisibility cloak with two corresponding "anti-objects" embedded in the complementary media layer.

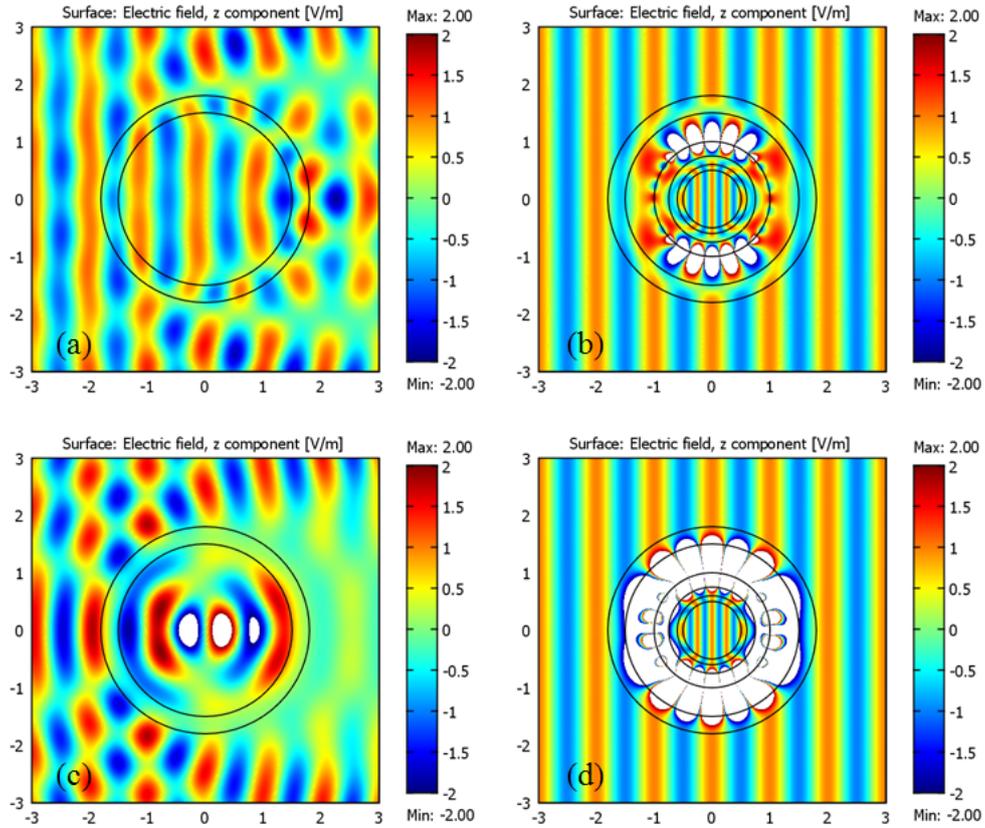

Fig. 3 (color online). Snapshots of the total electric fields under an incident TE plane wave from the left. (a) A dielectric circular shell with permittivity $\varepsilon_o = 2$. (b) The circular shell in (a) is cloaked by an invisibility cloak composed of a circular layer of complementary media with an embedded "anti-object" shell and a core material inside the shell. (c) A circular shell with permittivity $\varepsilon_o = -1$. (d) The circular shell in (c) is cloaked by a similar invisibility cloak of that in (b).